\documentclass[reprint,showpacs,preprintnumbers,amsmath,amssymb,prl]{revtex4-1}
\usepackage[hidelinks]{hyperref}
\usepackage{amsmath}
\usepackage{amsfonts}
\usepackage{amsthm}
\usepackage{dsfont}
\usepackage{graphics}
\usepackage{graphicx,bm}
\usepackage[caption=false]{subfig}
\usepackage{amssymb}
\usepackage{mathrsfs}
\usepackage{braket}
\usepackage{float}
\usepackage{color}
\usepackage{url}
\usepackage[abs]{overpic}
\usepackage[usenames,dvipsnames]{xcolor}
\usepackage{commath}
\usepackage{subfig}
\usepackage{multirow}
\usepackage{verbatim}
\usepackage{physics}
\usepackage{array}

\newcommand{\ii}{\mathrm{i}}
\newcommand{\ee}{\mathrm{e}}
\begin{document}

\title{A new method to building Dirac quantum walks coupled to electromagnetic fields}

\author{Gareth Jay$^{1}$}
\email{gareth.jay@uwa.edu.au}
\author{Fabrice Debbasch$^{2}$}
\email{fabrice.debbasch@gmail.com}
\author{J. B. Wang$^{1}$}
\email{jingbo.wang@uwa.edu.au}
\affiliation{{$^{1}$Physics Department, The University of Western Australia, Perth, WA 6009, Australia}\\
{$^{2}$Sorbonne Universit\'e, Observatoire de Paris, Universit\'e PSL, CNRS, LERMA, F-75005, {\sl Paris}, France}}

\date{\today}
\begin{abstract}
A quantum walk whose continuous limit coincides with Dirac equation is usually called a Dirac Quantum Walk (DQW). A new systematic method to build DQWs coupled to electromagnetic (EM) fields is 
introduced and put to test on several examples of increasing difficulty. It is first used to derive the EM coupling of a well-known $3D$ walk on the cubic lattice. Recently introduced DQWs on the triangular and honeycomb lattice are then re-derived, showing for the first time that these are the only DQWs that can be defined with spinors living on the vertices of these lattices. As a third example of the method's effectiveness, a new $3D$ walk on a parallelepiped lattice is derived. As a fourth, negative example, it is shown that certain lattices like the rhombohedral lattice cannot be used to build DQWs. The effect of changing representation in the Dirac equation is also discussed. 

\end{abstract}
%\pacs{03.67.-a, 47.37.+q, 47.40.-x, 67.10.-j}
\keywords{gggggg}
\maketitle

\section{Introduction}
Quantum walks are unitary quantum automata first proposed by Feynman \cite{feynman2010quantum,schweber1986feynman}, that can be viewed as formal generalizations of classical random walks. First introduced systematically by Aharonov et al. \cite{aharonov1993quantum} and Myers \cite{meyer1996quantum}, they have found application in quantum information and algorithmic development \cite{ambainis2007quantum,magniez2011search,ManouchehriWang2014}. They can also be used as quantum simulators \cite{Strauch2006, Strauch2007, Kurzynski2008, Chandrashekar2013, Shikano2013, Arrighi2014,arrighi2016quantum, Molfetta2014, perez2016asymptotic}, where the lattice represents a discretization of continuous space. More ambitiously, quantum walks may represent a potentially realistic discrete spacetime underlying the apparently continuous physical universe \cite{bisio2016special}.

It has been shown that several Discrete-Time Quantum Walks (DTQWs) defined on regular square lattices simulate the Dirac dynamics in various spacetime dimensions and that these Dirac Quantum Walks (DQWs) can be coupled to various discrete gauge fields \cite{di2013quantum,di2014quantum,arnault2016landau,arnault2016quantum,arnault2016quantum2,arnault2017quantum,bisio2015quantum,marquez2018electromagnetic, bialynicki1994weyl}. It has also more recently been shown that $2D$ DQWs can be defined on regular non-square lattices like the triangle and honeycomb lattice \cite{jay2018dirac,arrighi2018dirac}. The aim of the present article is to extend these results by developing a new systematic approach to construct DQWs coupled to EM fields. The new approach is based on discretizing the Hamiltonian \cite{arrighi2014dirac} in terms of directional derivatives \cite{arrighi2018dirac}. It also presupposes that the walk wave-function and gauge fields live on the vertices of the lattice and that, given a target space-time dimension $D$, the number of components of the wave function coincide with the number of components of the Dirac spinor in irreducible representations of the $D$-dimensional Lorentz group. This approach is superior to a trial-and-error method because (i) it shows unambiguously if a DQW can be constructed on a given lattice (ii) it delivers automatically the coefficients of the DQW (iii) it becomes necessary to use a systematic approach if one wants to deal with physical spaces of dimensions higher than $2$. 

We present the method  and put it to test on several examples of increasing difficulty. We first derive the EM coupling of a well-known $3D$ walk on the cubic lattice. We then re-derive recently introduced DQWs on the triangular and honeycomb lattice, showing for the first time that these are the only DQWs that can be defined with spinors living on the vertices of these lattices. As a third example of the method's effectiveness, we build a new $3D$ walk on a parallelepiped lattice and, as a fourth and negative example, we show that certain lattices like the rhombohedral lattice cannot be used to build DQWs. We finally discuss the effect of changing representation in the Dirac equation and mention possible extensions of this work. In an appendix we provide a step-by-step walkthrough of the method starting from the simplest $(1+1)D$ free Dirac equation, and working our way up through the dimensions and coupled fields.

\section{The Hamiltonian approach}

A general DTQW takes the form of
\begin{equation}\label{generalDTQW}
	\Psi(t+\Delta t)=\hat{W}\Psi(t),
\end{equation}
where the wavefunction $\Psi$ is an $N$-component spinor and $\hat{W}$ in a unitary operator belonging to $U(N)$, whereas the Dirac Equation, when written in so-called 'Hamiltonian form' takes a Schr{\"o}dinger-like equation form of
 \begin{equation}\label{DiracHamiltonian}
 	\ii\pdv{\Psi}{t}=\hat{H} \Psi.
 \end{equation}
If we discretize the time derivative, equation \ref{DiracHamiltonian} becomes
\begin{equation}
	\ii\left(\frac{\Psi(t+\epsilon)-\Psi(t)+\mathcal{O}(\epsilon^2)}{\epsilon}\right)=\hat{H} \Psi(t),
\end{equation}
which after some rearrangement becomes
\begin{equation}\label{rearrangedHamiltonianFormDirac}
	\Psi(t+\epsilon)=\left(\mathbb{I}_N-\ii\epsilon\hat{H}+\mathcal{O}(\epsilon^2)\right)\Psi(t),
\end{equation}
and so by matching equation \ref{generalDTQW} with \ref{rearrangedHamiltonianFormDirac}, we are then looking for an operator $\hat{W}$ that is of the form
\begin{equation}\label{defnW}
	\hat{W}=\mathbb{I}_N-\ii\epsilon\hat{H}+\mathcal{O}(\epsilon^2),
\end{equation}
where $N$ will be $4$ if our Dirac spinor is operating in $(3+1)D$ spacetime, or $2$ if in lower dimensions.

A quantum walk is a product of unitary operations, each representing either a translation or a mixing of states, acting on the wavefunction. So we need to factorize the expression in equation \ref{rearrangedHamiltonianFormDirac} to become a product of unitary operators. Fortunately due to the approximation to first order, any expression of the form
\begin{equation}\label{factorA}
	\mathbb{I}_N+\epsilon A+\epsilon B+\epsilon C+\mathcal{O}(\epsilon^2),
\end{equation}
can be factorized into
\begin{equation}\label{factorB}
	(\mathbb{I}_N+\epsilon A)(\mathbb{I}_N+\epsilon B)(\mathbb{I}_N+\epsilon C)+\mathcal{O}(\epsilon^2).
\end{equation}

For a step-by-step walkthrough, starting from the simplest case of a free Dirac equation in $(1+1)D$ flat spacetime, see the appendix. Here we shall present the more complicated case of a Dirac equation in $(3+1)D$ flat spacetime coupled to an electromagnetic field before moving onto non-cube lattices.

In $(3+1)D$ flat spacetime, the Hamiltonian $\hat{H}$ in equation \ref{DiracHamiltonian}, if the standard Dirac representation for the Dirac gamma matrices is chosen, is written as
\begin{equation}\label{H3DEM}
    \hat{H}=-\mathbb{I}_4A_0-\ii(\sigma_1\otimes\sigma_j)\partial_j-(\sigma_1\otimes\sigma_j)A_j+(\sigma_3\otimes\mathbb{I}_2)m.
\end{equation}
Plugging equation \ref{H3DEM} into equation \ref{defnW} we get
\begin{eqnarray}
    \hat{W}&=&\mathbb{I}_4+\ii\epsilon A_0\mathbb{I}_4-\sum_{j=1}^3\epsilon(\sigma_1\otimes\sigma_j)\partial_j+\sum_{j=1}^3\ii\epsilon A_j(\sigma_1\otimes\sigma_j)\nonumber\\&&-\ii m\epsilon(\sigma_3\otimes\mathbb{I}_2)+\mathcal{O}(\epsilon^2),
\end{eqnarray}
where if we make use of the factorizing trick from equations \ref{factorA} to \ref{factorB} we get
\begin{eqnarray}
    \hat{W}&=&(\mathbb{I}_4-\ii m\epsilon(\sigma_3\otimes\mathbb{I}_2))(\mathbb{I}_4+\ii\epsilon A_0\mathbb{I}_4)\nonumber\\&&\times\prod^3_{j=1}\left(\mathbb{I}_4+\ii\epsilon A_j(\sigma_1\otimes\sigma_j)\right)\prod^3_{j=1}\left(\mathbb{I}_4-\epsilon(\sigma_1\otimes\sigma_j)\partial_j\right)\nonumber\\&&+\mathcal{O}(\epsilon^2).\label{CubeWFactored}
\end{eqnarray}
The mass and electromagnetic field terms are easy enough to deal with, however the partial derivative terms require a bit more work since each shift operator can be described as
\begin{eqnarray}
    \hat{S}_j=e^{\epsilon(\sigma_3\otimes\sigma_3)\partial_j}=\mathbb{I}_4+\epsilon(\sigma_3\otimes\sigma_3)\partial_j+\mathcal{O}(\epsilon^2),
\end{eqnarray}
and so we need to factor out a unitary transformation of the form
\begin{equation}\label{UnitaryTransform}
    U_j\sigma_3U_j^\dagger=\sigma_j,
\end{equation}
which is described in more detail in the appendix. The partial derivative terms can then be expressed as
\begin{equation}
    \mathbb{I}_4-\epsilon(\sigma_1\otimes\sigma_j)\partial_j=\\(U_1\sigma_1\otimes U_j)\hat{S}_j(\sigma_1U_1^\dagger\otimes U_j^\dagger).
\end{equation}
Finally making use of the expansions \ref{cos}, \ref{sin} and \ref{exp} from the appendix you find the operator $\hat{W}$ to be
\begin{eqnarray}
    \hat{W}&=&\begin{pmatrix}
        \ee^{-\ii m\epsilon}\mathbb{I}_2&0\\
        0&\ee^{\ii m\epsilon}\mathbb{I}_2
    \end{pmatrix}\ee^{\ii\epsilon A_0}\nonumber\\&&\times\prod^3_{j=1}\begin{pmatrix}\cos{(\epsilon A_j)}\mathbb{I}_2&\ii\sin{(\epsilon A_j)}\sigma_j\\\ii\sin{(\epsilon A_j)}\sigma_j&\cos{(\epsilon A_j)}\mathbb{I}_2\end{pmatrix}\nonumber\\&&\times\prod^3_{j=1}\left((U_1\sigma_1\otimes U_j)\hat{S}_j(\sigma_1U_1^\dagger\otimes U_j^\dagger)\right).\label{CubeWFinal}
\end{eqnarray}
So the DTQW that is defined by this operator $\hat{W}$ will in its continuous limit coincides with the $(3+1)D$ Dirac equation couples to an electromagnetic field $A_\mu$. We shall now extend this procedure to non-cube lattices.

\section{Non-Cube Lattices}
In the attempt to extend this procedure to non-cube lattices we shall start in the simpler $(2+1)D$ spacetime and derive, from the Hamiltonian, two of the triangular lattice DQWs found in Jay et al. \cite{jay2018dirac}.

When looking at non-cube lattices we must make use of directional derivatives and rewrite the Dirac equation in terms of these. This will often bring in coefficients that make the quantum walk incompatible with the Dirac equation if you insist on keeping the lattice step size the same in each spacetime direction. This can be seen clearly in the DQW on a triangular lattice.

\subsection{Triangular Lattice}

\begin{figure}[t]
\includegraphics[width=.65\linewidth]{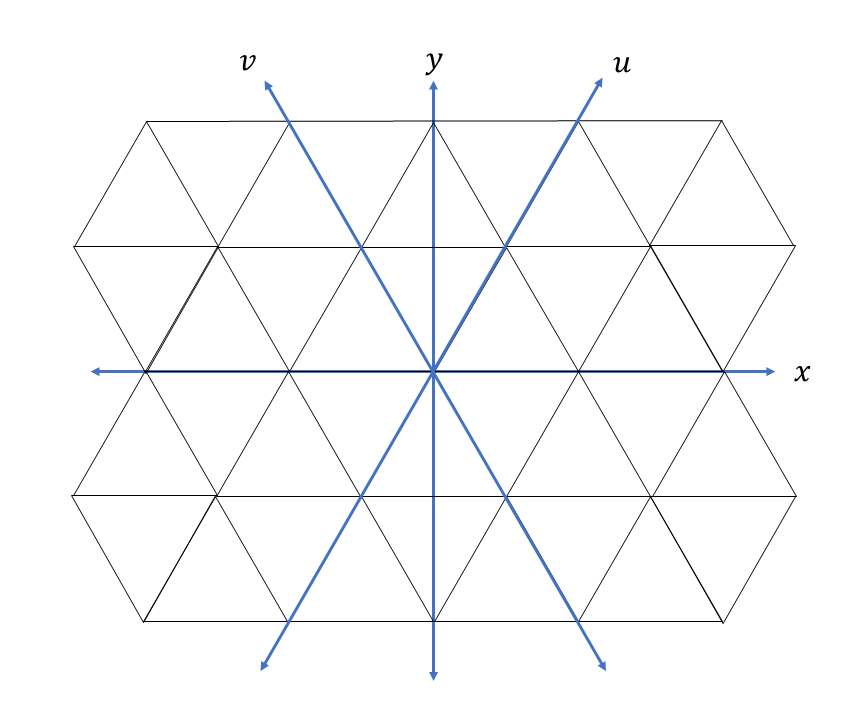}
\caption{Equilateral Triangular lattice embedded on a standard euclidean 2D space showing the two new directions $u$ and $v$}
\label{triangularlattice}
\end{figure}

Consider the equilateral triangular lattice in Fig. \ref{triangularlattice}. We can define two unit vectors in the direction of $u$ and $v$ as
\begin{eqnarray*}
    \hat{u}&=&\left(\frac{1}{2},\frac{\sqrt{3}}{2}\right),\\
    \hat{v}&=&\left(-\frac{1}{2},\frac{\sqrt{3}}{2}\right).
\end{eqnarray*}
These two unit vectors define the two directions that make this lattice different to the square lattices. (Our third direction across this lattice will still be the same as the $x$-direction in the square lattice).
The directional derivatives associated with these directions are then
\begin{eqnarray*}
    \frac{\partial}{\partial u}=\vec{\nabla}\cdot\hat{u}&=&\frac{1}{2}\frac{\partial}{\partial x}+\frac{\sqrt{3}}{2}\frac{\partial}{\partial y},\\
    \frac{\partial}{\partial v}=\vec{\nabla}\cdot\hat{v}&=&-\frac{1}{2}\frac{\partial}{\partial x}+\frac{\sqrt{3}}{2}\frac{\partial}{\partial y}.\\
\end{eqnarray*}
Summing the two definitions we get the relationship
\begin{equation}
    \frac{\partial}{\partial y}=\frac{1}{\sqrt{3}}\left(\frac{\partial}{\partial u}+\frac{\partial}{\partial v}\right).\label{yDerivTri}
\end{equation}
The new directional shift operators are going to behave very similar to the standard $x-$ and $y-$direction ones, as can be seen when we define them:

\begin{eqnarray}
    \hat{S}_u\Psi(x,y,t)&=&\begin{pmatrix}
                        \psi_L(x+\frac{\epsilon}{2},y+\frac{\sqrt{3}\epsilon}{2},t)\\
                        \psi_R(x-\frac{\epsilon}{2},y-\frac{\sqrt{3}\epsilon}{2},t)
                            \end{pmatrix}\nonumber\\
                    &=&\begin{pmatrix}
                 		\psi_L(x,y,t)\\
                        \psi_R(x,y,t)
                    \end{pmatrix}+\epsilon\sigma_3\left(\frac{1}{2}\partial_1+\frac{\sqrt{3}}{2}\partial_2\right)\nonumber\\&&\times\begin{pmatrix}
                 												\psi_L(x,y,t)\\
                       											\psi_R(x,y,t)
                    										\end{pmatrix}+\mathcal{O}(\epsilon^2)\nonumber\\
                 &=&\begin{pmatrix}
                 		\psi_L(x,y,t)\\
                        \psi_R(x,y,t)
                    \end{pmatrix}+\epsilon\sigma_3\partial_u\begin{pmatrix}
                 												\psi_L(x,y,t)\\
                       											\psi_R(x,y,t)
                    										\end{pmatrix}\nonumber\\&&+\mathcal{O}(\epsilon^2)\nonumber\\
                 &=&\left(\mathbb{I}_2+\epsilon\sigma_3\partial_u\right)\begin{pmatrix}
                 												\psi_L(x,y,t)\\
                       											\psi_R(x,y,t)
                    										\end{pmatrix}+\mathcal{O}(\epsilon^2)\nonumber\\
              &=&\ee^{\epsilon\sigma_3\partial_u}\Psi(x,y,t).
\end{eqnarray}
$\hat{S}_v$ is defined the exact same way and note since we are in the lower $(2+1)$ dimensions, $\hat{S}_1$ for the $x-$direction would also fit this pattern (see appendix). If we now plug equation \ref{yDerivTri} into the $(2+1)D$ free Dirac Equation's Hamiltonian we get 
\begin{eqnarray}
     \hat{H}&=&\ii\sigma_3\partial_1-\ii\sigma_2\partial_2+\sigma_1 m\nonumber\\
     &=&\ii\sigma_3\partial_1-\ii\sigma_2\frac{1}{\sqrt{3}}\left(\partial_u+\partial_v\right)+\sigma_1 m.
\end{eqnarray}
Plugging this into our operator $\hat{W}$ we get
\begin{eqnarray}
     \hat{W}&=&\mathbb{I}_2-\ii\epsilon\hat{H}+\mathcal{O}(\epsilon^2)\nonumber\\
     &=&\mathbb{I}_2+\epsilon\sigma_3\partial_1-\epsilon\sigma_2\frac{1}{\sqrt{3}}\partial_u-\epsilon\sigma_2\frac{1}{\sqrt{3}}\partial_v-\ii m\epsilon\sigma_1+\mathcal{O}(\epsilon^2)\nonumber\\
     &=&(\mathbb{I}_2-\ii m\epsilon\sigma_1)\left(1-\epsilon\sigma_2\frac{1}{\sqrt{3}}\partial_v\right)\left(1-\epsilon\sigma_2\frac{1}{\sqrt{3}}\partial_u\right)\nonumber\\&&\times(\mathbb{I}_2+\epsilon\sigma_3\partial_1)+\mathcal{O}(\epsilon^2).\nonumber\\
\end{eqnarray}
Now if we were to take the approach we use in the square lattice we would be looking for a unitary operators $U$ that follows the relationship
\begin{eqnarray}
     U\sigma_3U^\dagger&=&-\frac{1}{\sqrt{3}}\sigma_2.
\end{eqnarray}
Unfortunately this relationship has no solution. There are two places we can go at this point. One involves changing the step size of the spatial directions, the other involves changing the step size in the time dimension.
\subsubsection{Isosceles Lattice (changing the $y$ step size)}

\begin{figure}[t]
\includegraphics[width=.65\linewidth]{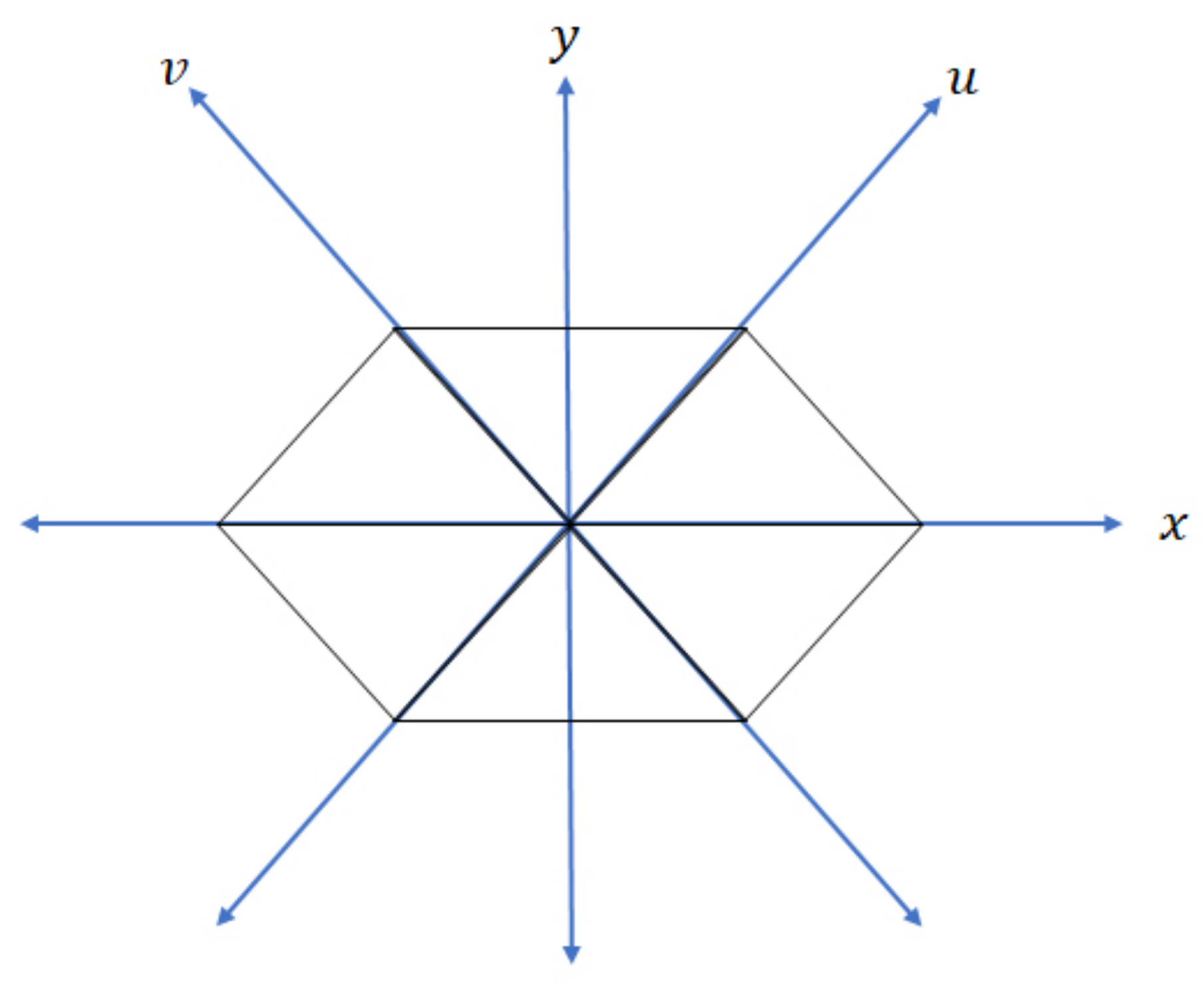}
\caption{Isosceles Triangular lattice embedded on a standard euclidean 2D space showing the two new directions $u$ and $v$}
\label{isolattice}
\end{figure}

The first method we shall consider is changing the step size of the spatial direction. We shall keep the $x-$direction at step size $\epsilon$ but will change the step size of the $y-$direction (effectively changing our $u$ and $v$ directions). This will effectively make our triangular lattice an isoceles lattice (see Fig. \ref{isolattice}). 

In our previous work \cite{jay2018dirac} we found a quantum walk that works on an isosceles lattice where the triangle height was $\frac{\epsilon}{2}$ which is effectively dilating the $y-$direction by a factor of $\frac{1}{\sqrt{3}}$. We shall however not start with this assumption, but place an arbitrary factor of $\upsilon$ with the hope of proving that we want $\upsilon=\frac{1}{\sqrt{3}}$.
We can now define our two unit vectors $\hat{u}$ and $\hat{v}$ as
\begin{eqnarray*}
    \hat{u}&=&\frac{2}{\sqrt{1+3\upsilon^2}}\left(\frac{1}{2},\frac{\upsilon\sqrt{3}}{2}\right),\\
    \hat{v}&=&\frac{2}{\sqrt{1+3\upsilon^2}}\left(-\frac{1}{2},\frac{\upsilon\sqrt{3}}{2}\right).
\end{eqnarray*}
Where the factor out the front is the normalisation factor for the unit vector that we can refer to as $N$. The directional derivatives associated with these directions are then
\begin{eqnarray*}
    \partial_u&=&N\left(\frac{1}{2}\partial_1+\frac{\upsilon\sqrt{3}}{2}\partial_2\right),\\
    \partial_v&=&N\left(-\frac{1}{2}\partial_1+\frac{\upsilon\sqrt{3}}{2}\partial_2\right).
\end{eqnarray*}
Summing these two definitions we get the relationship
\begin{eqnarray}
    \partial_2=\frac{1}{N\upsilon\sqrt{3}}\left(\partial_u+\partial_v\right).\label{triangleDerivates}
\end{eqnarray}
The new directional shift operators are going to behave very similar but the normalisation constant $N$ will turn up
\begin{eqnarray}
    \hat{S}_u\Psi(x,y,t)&=&\begin{pmatrix}
                        \psi_L(x+\frac{\epsilon}{2},y+\frac{\sqrt{3}\epsilon}{2},t)\\
                        \psi_R(x-\frac{\epsilon}{2},y-\frac{\sqrt{3}\epsilon}{2},t)
                            \end{pmatrix}\nonumber\\
                    &=&\begin{pmatrix}
                 		\psi_L(x,y,t)\\
                        \psi_R(x,y,t)
                    \end{pmatrix}\nonumber\\&&+\epsilon\sigma_3(\frac{1}{2}\partial_1+\frac{\sqrt{3}}{2}\partial_2)\begin{pmatrix}
                 												\psi_L(x,y,t)\\
                       											\psi_R(x,y,t)
                    										\end{pmatrix}+\mathcal{O}(\epsilon^2)\nonumber\\
                &=&\begin{pmatrix}
                 		\psi_L(x,y,t)\\
                        \psi_R(x,y,t)
                    \end{pmatrix}\nonumber\\&&+\epsilon\sigma_3\frac{1}{N}\partial_u\begin{pmatrix}
                 												\psi_L(x,y,t)\\
                       											\psi_R(x,y,t)
                    										\end{pmatrix}+\mathcal{O}(\epsilon^2)\nonumber\\
                 &=&\left(\mathbb{I}_2+\epsilon\sigma_3\frac{1}{N}\partial_u\right)\begin{pmatrix}
                 												\psi_L(x,y,t)\\
                       											\psi_R(x,y,t)
                    										\end{pmatrix}+\mathcal{O}(\epsilon^2)\nonumber\\
              &=&\ee^{\frac{1}{N}\epsilon\sigma_3\partial_u}\Psi(x,y,t).
\end{eqnarray}
$\hat{S}_v$ is defined the exact same way, while $\hat{S}_1$ will be completely unaffected by the dilation of the $y-$direction. You can see $\hat{S}_u$ and $\hat{S}_v$ appear very similar to before however the $\frac{1}{N}$ factor is appearing. The Hamiltonian now becomes
\begin{eqnarray}
    \hat{H}&=&\ii\sigma_3\partial_1-\ii\sigma_2\partial_2+\sigma_1m\nonumber\\
   &=&\ii\sigma_3\partial_1-\ii\sigma_2\frac{1}{N\upsilon\sqrt{3}}(\partial_u+\partial_v)+\sigma_1m.
\end{eqnarray}
The factor attached to our $\partial_u$ and $\partial_v$ terms is now $\frac{1}{N\upsilon\sqrt{3}}$. The $N$ part of this is not a problem as we want the $\frac{1}{N}$ factor for our shift operators this time round. The problem is the $\upsilon\sqrt{3}$ part. This is where is becomes clear that the choice of $\upsilon$ should indeed be $\frac{1}{\sqrt{3}}$ as seen in the previous work \cite{jay2018dirac}. This will lead to the $\hat{W}$ operator being
\begin{eqnarray}
    \hat{W}&=&\mathbb{I}_2-\ii\epsilon\hat{H}+\mathcal{O}(\epsilon^2)\nonumber\\
    &=&\mathbb{I}_2-\ii m\epsilon\sigma_1-\frac{1}{N}\epsilon\sigma_2\partial_v-\frac{1}{N}\epsilon\sigma_2\partial_u+\epsilon\sigma_3\partial_1+\mathcal{O}(\epsilon^2)\nonumber\\
    &=&(\mathbb{I}_2-\ii m\epsilon\sigma_1)(\mathbb{I}_2-\frac{1}{N}\epsilon\sigma_2\partial_v)(\mathbb{I}_2-\frac{1}{N}\epsilon\sigma_2\partial_u)\nonumber\\&&\times(\mathbb{I}_2+\epsilon\sigma_3\partial_1)+\mathcal{O}(\epsilon^2)\nonumber\\
    &=&(\mathbb{I}_2-\ii m\epsilon\sigma_1)U(\mathbb{I}_2+\frac{1}{N}\epsilon \sigma_3\partial_v)U^\dagger \nonumber\\&&\times U(\mathbb{I}_2+\frac{1}{\sqrt{2}}\epsilon \sigma_3\partial_u)U^\dagger(\mathbb{I}_2+\epsilon\sigma_3\partial_1)+\mathcal{O}(\epsilon^2)\nonumber\\
    &=&(\mathbb{I}_2-\ii m\epsilon\sigma_1)U\hat{S}_vU^\dagger U\hat{S}_uU^\dagger\hat{S}_1+\mathcal{O}(\epsilon^2)\nonumber\\
    &=&\begin{pmatrix}
            \cos(m\epsilon)&-\ii\sin(m\epsilon)\\
            -\ii\sin(m\epsilon)&cos(m\epsilon)
        \end{pmatrix}U\hat{S}_v\hat{S}_uU^\dagger\hat{S}_1,
\end{eqnarray}
where $U$ is a unitary matrix such that
\begin{equation}
	U\sigma_3U^\dagger=-\sigma_2,
\end{equation}
thus $U$ follows the same rules as described in the square lattice case in the appendix. The $U$ we shall choose is
\begin{equation}
U=\begin{pmatrix}
\frac{1}{\sqrt{2}}&-\frac{i}{\sqrt{2}}\\
-\frac{i}{\sqrt{2}}&\frac{1}{\sqrt{2}}
\end{pmatrix}.
\end{equation}

Note that this operator is the same as the operator for the square lattice, except that $\hat{S}_2$ is replaced with the composition $\hat{S}_v\hat{S}_u$. This is directly related to the relationship in equation \ref{triangleDerivates}. This lines up perfectly with the isosceles walk defined in our previous paper \cite{jay2018dirac}. The other method of making a DQW work on a triangular lattice is by changing the time step.

\subsubsection{Equilateral Triangle Lattice (changing the $t$ step size)}

From previous work we have found that unlike the isosceles triangular lattice, walks across equilateral triangles needed the step size in time to be different to that in the spatial directions \cite{jay2018dirac}. So this is the place to start with the other method to keep the lattice equilateral. When we set 
\begin{equation*}
    \hat{W}=\mathbb{I}_2-\ii\epsilon\hat{H}+\mathcal{O}(\epsilon^2)
\end{equation*}
at the beginning, we are implicitly setting the step size in time to be $\Delta t=\epsilon$. If instead we make this an arbitrary factor of $\epsilon$, i.e. $\Delta t=\alpha \epsilon$ and feed this into the definition for $\hat{W}$ we get
\begin{equation}
    \hat{W}=\mathbb{I}_2-\ii\alpha\epsilon\hat{H}+\mathcal{O}(\epsilon^2).
\end{equation}
Subsequently feed this into what we had before and we get 
\begin{equation*}
    \hat{W}=\mathbb{I}_2-\ii m\alpha\epsilon\sigma_1-\frac{1}{\sqrt{3}}\alpha\epsilon\sigma_2\partial_v-\frac{1}{\sqrt{3}}\alpha\epsilon\sigma_2\partial_u+\alpha\epsilon\sigma_3\partial_1+\mathcal{O}(\epsilon^2).
\end{equation*}
Now one would be tempted to choose $\alpha=\sqrt{3}$ which would reduce the middle terms to what we had before and we're good to go, but of course there are other $\alpha$ terms hanging around now. It is not essentially a problem in the mass term, however in the $\partial_1$ term, we now have a problem we did not have before. 
A work around for this is if we split this term up into two parts of size $\frac{1}{\alpha}$ and $\frac{\alpha-1}{\alpha}$ we get
\begin{eqnarray}
     \hat{W}&=&\mathbb{I}_2-\ii m\alpha\epsilon\sigma_1-\frac{1}{\sqrt{3}}\alpha\epsilon\sigma_2\partial_v-\frac{1}{\sqrt{3}}\alpha\epsilon\sigma_2\partial_u\nonumber\\&&+\alpha\left(\frac{1}{
     \alpha}+\frac{\alpha-1}{\alpha}\right)\epsilon\sigma_3\partial_1+\mathcal{O}(\epsilon^2)\nonumber\\
    &=&\mathbb{I}_2-\ii m\alpha\epsilon\sigma_1-\frac{1}{\sqrt{3}}\alpha\epsilon\sigma_2\partial_v-\frac{1}{\sqrt{3}}\alpha\epsilon\sigma_2\partial_u\nonumber\\&&+\left(\alpha-1\right)\epsilon\sigma_3\partial_1+\epsilon\sigma_3\partial_1+\mathcal{O}(\epsilon^2).\nonumber\\
\end{eqnarray}
This essentially gives us an $\epsilon\sigma_3\partial_1$ term that will work normally and head towards an $\hat{S}_1$ operator, and a more troublesome $\alpha-1$ term. This however can be moved into the $\partial_u$ and $\partial_v$ terms by making a relationship for $\partial_1$ through the subtraction of the directional derivatives:
\begin{equation}
    \partial_1=\partial_u-\partial_v.
\end{equation}
Subbing these in we get
\begin{eqnarray}
    \hat{W}&=&\mathbb{I}_2-\ii m\alpha\epsilon\sigma_1-\frac{1}{\sqrt{3}}\alpha\epsilon\sigma_2\partial_v-\frac{1}{\sqrt{3}}\alpha\epsilon\sigma_2\partial_u\nonumber\\&&+\left(\alpha-1\right)\epsilon\sigma_3(\partial_u-\partial_v)+\epsilon\sigma_3\partial_1+\mathcal{O}(\epsilon^2)\nonumber\\ 
    &=&\mathbb{I}_2-\ii m\alpha\epsilon\sigma_1-\left(\left(\alpha-1\right)\sigma_3+\frac{\alpha}{\sqrt{3}}\sigma_2\right)\epsilon\partial_v\nonumber\\&&+\left(\left(\alpha-1\right)\sigma_3-\frac{\alpha}{\sqrt{3}}\sigma_2\right)\epsilon\partial_u+\epsilon\sigma_3\partial_1+\mathcal{O}(\epsilon^2)\nonumber\\ 
    &=&(\mathbb{I}_2-\ii m\alpha\epsilon\sigma_1)\left(\mathbb{I}_2-\left(\left(\alpha-1\right)\sigma_3+\frac{\alpha}{\sqrt{3}}\sigma_2\right)\epsilon\partial_v\right)\nonumber\\&&\times\left(\mathbb{I}_2+\left(\left(\alpha-1\right)\sigma_3-\frac{\alpha}{\sqrt{3}}\sigma_2\right)\epsilon\partial_u\right)(\mathbb{I}_2+\epsilon\sigma_3\partial_1)\nonumber\\&&+\mathcal{O}(\epsilon^2).
\end{eqnarray}
So now we're looking for two unitary operators $U_v$ and $U_u$ such that
\begin{eqnarray}
    U_v\sigma_3U_v^\dagger&=&-\left(\left(\alpha-1\right)\sigma_3+\frac{\alpha}{\sqrt{3}}\sigma_2\right),\nonumber\\
    U_u\sigma_3U_u^\dagger&=&\left(\left(\alpha-1\right)\sigma_3-\frac{\alpha}{\sqrt{3}}\sigma_2\right),
\end{eqnarray}
where $\alpha\ne1$ and $\alpha\ne0$.
Again these unitary operators will not be unique, however we will see that the choice of $\alpha$ is surprisingly forced upon us. Let us consider the first relationship above and parameterise $U_v$ as
\begin{equation*}
    U_v=\begin{pmatrix}a&b\\c&d\end{pmatrix}.
\end{equation*}
The result follows that
\begin{eqnarray}
    |a|^2=|d|^2&=&1-\frac{\alpha}{2},\nonumber\\
    |c|^2=|b|^2&=&\frac{\alpha}{2},\nonumber\\
    a\bar{c}=d\bar{b}&=&\frac{\ii\alpha}{2\sqrt{3}}.
\end{eqnarray}
Now if we parameterise each of the four components in the style of $a=|a|\ee^{\ii\theta_a}$ and rewrite the $a\bar{c}$ part of the third equation above, (this works similarly for $b$ and $d$),
\begin{eqnarray}
    |a|\ee^{\ii\theta_a}|c|\ee^{-\ii\theta_c}&=&\frac{\ii\alpha}{2\sqrt{3}}\nonumber\\
    |a||c|\ee^{\ii(\theta_a-\theta_c)}&=&\frac{\ii}{2\sqrt{3}}\alpha\nonumber\\
    \sqrt{1-\frac{\alpha}{2}}\sqrt{\frac{\alpha}{2}}\ee^{\ii(\theta_a-\theta_c)}&=&\frac{\ii}{2\sqrt{3}}\alpha\nonumber\\
    \frac{\alpha}{2}-\frac{\alpha^2}{4}&=&-\frac{1}{12}\ee^{\ii(\theta_c-\theta_a)}\alpha^2\nonumber\\
    \frac{6}{\alpha}-3&=&-\ee^{\ii(\theta_c-\theta_a)}\nonumber\\
    \frac{6}{\alpha}&=&3-\ee^{\ii(\theta_c-\theta_a)}\nonumber\\
    \alpha&=&\frac{6}{3-\ee^{\ii(\theta_c-\theta_a)}}.\nonumber\\
\end{eqnarray}
Now it is good to remember here that $\alpha$ represents a step in time, and so must be real, not complex. This means that the $\ee^{\ii(\theta_c-\theta_a)}$ term must equal $\pm1$. This would result in $\alpha$ being either $3$ or $\frac{3}{2}$. However if $\alpha=3$ then with the above relations $|a|^2=|d|^2=-\frac{1}{2}$ which can't be true since modulas squared values need to be positive. And so we show $\alpha$ must equal $\frac{3}{2}$. Once this is determined we have the relations
\begin{eqnarray}
    |a|^2=|d|^2&=&\frac{1}{4},\nonumber\\
    |c|^2=|b|^2&=&\frac{3}{4},\nonumber\\
    a\bar{c}=d\bar{b}&=&\frac{\ii\sqrt{3}}{4}.
\end{eqnarray}
We now have two degrees of freedom so if we choose the simplest choice for $c$ and $b$ to be $\frac{\sqrt{3}}{2}$ we get
\begin{equation}
    U_v=\begin{pmatrix}\frac{\ii}{2}&\frac{\sqrt{3}}{2}\\
    \frac{\sqrt{3}}{2}&\frac{\ii}{2}\end{pmatrix}.
\end{equation}
Through a similar technique you can determine $U_u$ as
\begin{equation}
    U_u=\begin{pmatrix}\frac{\sqrt{3}}{2}&-\frac{\ii}{2}\\
    -\frac{\ii}{2}&\frac{\sqrt{3}}{2}\end{pmatrix}.
\end{equation}
We then continue as before with the knowledge of $\alpha$ and our unitary matrices
\begin{eqnarray}
    \hat{W}&=&(\mathbb{I}_2-im\frac{3}{2}\epsilon\sigma_1)\left(\mathbb{I}_2-\left(\frac{1}{2}\sigma_3+\frac{\sqrt{3}}{2}\sigma_2\right)\epsilon\partial_v\right)\nonumber\\&&\times\left(\mathbb{I}_2+\left(\frac{3}{2}\sigma_3-\frac{\sqrt{3}}{2}\sigma_2\right)\epsilon\partial_u\right)(\mathbb{I}_2+\epsilon\sigma_3\partial_1)+\mathcal{O}(\epsilon^2)\nonumber\\
    &=&(\mathbb{I}_2-im\frac{3}{2}\epsilon\sigma_1)\left(\mathbb{I}_2+U_v\sigma_3U_v^\dagger\epsilon\partial_v\right)\nonumber\\&&\times\left(\mathbb{I}_2+U_u\sigma_3U_u^\dagger\epsilon\partial_u\right)\hat{S}_1+\mathcal{O}(\epsilon^2)\nonumber\\
    &=&\begin{pmatrix}\cos(\frac{3}{2}m\epsilon)&-i\sin(\frac{3}{2}m\epsilon)\\-i\sin(\frac{3}{2}m\epsilon)&\cos(\frac{3}{2}m\epsilon)\end{pmatrix}U_v\hat{S}_vU_v^\dagger U_u\hat{S}_uU_u^\dagger\hat{S}_1.\nonumber\\
\end{eqnarray}

This matches the equilateral walk found in previous work \cite{jay2018dirac} and also proves that the $\frac{3}{2}$ time step is a required choice. Note however that the $U$ operators found here are \emph{slightly} different choice to the ones found in \cite{jay2018dirac} due to the fact that the choice of $\gamma$ matrices are different. We have now seen enough to extend this to 3D lattices.

\subsection{Parallelepiped Lattice}

\begin{figure}[b]
\includegraphics[width=.65\linewidth]{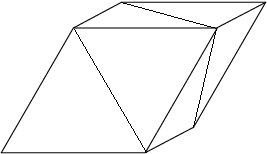}
\caption{A rhombohedron}
\label{rhombohedron}
\end{figure}

An obvious question about the non-square lattices where our last paper left off \cite{jay2018dirac} is whether or not the idea can be extended to $(3+1)D$ spacetime. So we shall attempt to apply the techniques we've discovered so far to a parallelepiped lattice. Consider the rhombohedron in Figure \ref{rhombohedron}. If we have the extra edges on the smallest diagonal of each rhombus face then we obtain a 3D lattice where on 3 different planes we can see the same triangular lattice as above. If the sides of the rhombohedron are of size $\epsilon$ then at any particular vertex of the lattice made up of this shape we would have the ability to move in the directions (or opposite directions) of the following unit vectors

\begin{eqnarray}
    \hat{x}&=&\left(1,0,0\right),\nonumber\\
    \hat{a}&=&\left(\frac{1}{2},\frac{1}{2\sqrt{3}},\sqrt{\frac{2}{3}}\right),\nonumber\\
    \hat{b}&=&\left(-\frac{1}{2},\frac{1}{2\sqrt{3}},\sqrt{\frac{2}{3}}\right),\nonumber\\
    \hat{c}&=&\left(\frac{1}{2},\frac{\sqrt{3}}{2},0\right),\nonumber\\
    \hat{d}&=&\left(-\frac{1}{2},\frac{\sqrt{3}}{2},0\right),\nonumber\\
    \hat{e}&=&\left(0,-\frac{1}{\sqrt{3}},\sqrt{\frac{2}{3}}\right).
\end{eqnarray}

Clearly we are going to run into the same issues with all those surds floating around as before, so we shall pre-empt these problems and again consider our two options - changing the spatial direction step size or changing the time step size.

\subsubsection{Parallelepiped Lattice (changing the $y$ and $z$ step sizes)}

We will introduce factors of $\upsilon$ and $\zeta$ this time to dilate the $y-$ and $z-$ directions respectively. This will of course mean we are no longer technically on a rhombohedral lattice but a more general parallelepiped lattice. Our unit vectors now become

\begin{eqnarray}
    \hat{x}&=&\left(1,0,0\right),\nonumber\\
    \hat{a}&=&\frac{2\sqrt{3}}{\sqrt{3+\upsilon^2+8\zeta^2}}\left(\frac{1}{2},\frac{\upsilon}{2\sqrt{3}},\zeta\sqrt{\frac{2}{3}}\right),\nonumber\\
    \hat{b}&=&\frac{2\sqrt{3}}{\sqrt{3+\upsilon^2+8\zeta^2}}\left(-\frac{1}{2},\frac{\upsilon}{2\sqrt{3}},\zeta\sqrt{\frac{2}{3}}\right),\nonumber\\
    \hat{c}&=&\frac{2}{\sqrt{1+3\upsilon^2}}\left(\frac{1}{2},\frac{\upsilon\sqrt{3}}{2},0\right),\nonumber\\
    \hat{d}&=&\frac{2}{\sqrt{1+3\upsilon^2}}\left(-\frac{1}{2},\frac{\upsilon\sqrt{3}}{2},0\right),\nonumber\\
    \hat{e}&=&\frac{\sqrt{3}}{\sqrt{\upsilon^2+2\zeta^2}}\left(0,-\frac{\upsilon}{\sqrt{3}},\zeta\sqrt{\frac{2}{3}}\right).
\end{eqnarray}

Making our corresponding directional derivatives

\begin{eqnarray}
    \partial_a&=&A\left(\frac{1}{2}\partial_1+\frac{\upsilon}{2\sqrt{3}}\partial_2+\zeta\sqrt{\frac{2}{3}}\partial_3)\right),\nonumber\\
    \partial_b&=&B\left(-\frac{1}{2}\partial_1+\frac{\upsilon}{2\sqrt{3}}\partial_2+\zeta\sqrt{\frac{2}{3}}\partial_3\right),\nonumber\\
    \partial_c&=&C\left(\frac{1}{2}\partial_1+\frac{\upsilon\sqrt{3}}{2}\partial_2\right),\nonumber\\
    \partial_d&=&D\left(-\frac{1}{2}\partial_1+\frac{\upsilon\sqrt{3}}{2}\partial_2\right),\nonumber\\
    \partial_e&=&E\left(-\frac{\upsilon}{\sqrt{3}}\partial_2+\zeta\sqrt{\frac{2}{3}}\partial_3\right),
\end{eqnarray}
where $A$, $B$, $C$, $D$ and $E$ are the normalization constants at the front of each of the unit vectors. Then recalling $A=B$ and $C=D$ we have the relationships
\begin{eqnarray}
     \partial_2&=&\frac{1}{\upsilon\sqrt{3}C}\left(\partial_c+\partial_d\right),\nonumber\\
     \partial_3&=&\frac{1}{\zeta\sqrt{6}}\left(\frac{1}{A}\left(\partial_a+\partial_b\right)+\frac{1}{E}\partial_e\right),
\end{eqnarray}
and the corresponding shift operators:
\begin{eqnarray}
     \hat{S}_a\Psi(x,y,z,t)&=&\begin{pmatrix}
                                   \psi_1(x+\frac{\epsilon}{2},y+\frac{\epsilon\upsilon}{2\sqrt{3}},z+\epsilon\zeta\sqrt{\frac{2}{3}},t)\\
                                   \psi_2(x-\frac{\epsilon}{2},y-\frac{\epsilon\upsilon}{2\sqrt{3}},z-\epsilon\zeta\sqrt{\frac{2}{3}},t)\\
                                   \psi_3(x-\frac{\epsilon}{2},y-\frac{\epsilon\upsilon}{2\sqrt{3}},z-\epsilon\zeta\sqrt{\frac{2}{3}},t)\\
                                   \psi_4(x+\frac{\epsilon}{2},y+\frac{\epsilon\upsilon}{2\sqrt{3}},z+\epsilon\zeta\sqrt{\frac{2}{3}},t)\\
                              \end{pmatrix}\nonumber\\
                        &=&\ee^{\frac{1}{A}\epsilon(\sigma_3\otimes\sigma_3)\partial_a}\Psi(x,y,z,t).
\end{eqnarray}
Each of the other shift operators follow this same pattern. Jumping to the Hamiltonian now we get
\begin{eqnarray}
    \hat{H}&=&(\sigma_3\otimes\mathbb{I}_2)m-\ii(\sigma_1\otimes\sigma_1)\partial_1-\ii(\sigma_1\otimes\sigma_2)\partial_2-\ii(\sigma_1\otimes\sigma_3)\partial_3\nonumber\\
    &=&(\sigma_3\otimes\mathbb{I}_2)m-\ii(\sigma_1\otimes\sigma_1)\partial_1-\ii(\sigma_1\otimes\sigma_2)\frac{1}{\upsilon\sqrt{3}C}\left(\partial_c+\partial_d\right)\nonumber\\&&-\ii(\sigma_1\otimes\sigma_3)\frac{1}{\zeta\sqrt{6}}\left(\frac{1}{A}\left(\partial_a+\partial_b\right)+\frac{1}{E}\partial_e\right).\nonumber\\
\end{eqnarray}
Now again remembering $A=B$ and $C=D$ then this becomes
\begin{eqnarray}
     \hat{H}&=&(\sigma_3\otimes\mathbb{I}_2)m-\ii(\sigma_1\otimes\sigma_1)\partial_1-\ii(\sigma_1\otimes\sigma_2)\frac{1}{C\upsilon\sqrt{3}}\partial_c\nonumber\\&&-\ii(\sigma_1\otimes\sigma_2)\frac{1}{D\upsilon\sqrt{3}}\partial_d-\ii(\sigma_1\otimes\sigma_3)\frac{1}{\zeta\sqrt{6}A}\partial_a\nonumber\\&&-\ii(\sigma_1\otimes\sigma_3)\frac{1}{\zeta\sqrt{6}B}\partial_b-\ii(\sigma_1\otimes\sigma_3)\frac{1}{\zeta\sqrt{6}E}\partial_e.
\end{eqnarray}
This certainly suggests sensible choices for $\upsilon$ and $\zeta$ are $\frac{1}{\sqrt{3}}$ and $\frac{1}{\sqrt{6}}$ respectively since this would give
\begin{eqnarray}
     \hat{H}&=&(\sigma_3\otimes\mathbb{I}_2)m-\ii(\sigma_1\otimes\sigma_1)\partial_1-\ii(\sigma_1\otimes\sigma_2)\frac{1}{C}\partial_c\nonumber\\&&-\ii(\sigma_1\otimes\sigma_2)\frac{1}{D}\partial_d-\ii(\sigma_1\otimes\sigma_3)\frac{1}{A}\partial_a\nonumber\\&&-\ii(\sigma_1\otimes\sigma_3)\frac{1}{B}\partial_b-\ii(\sigma_1\otimes\sigma_3)\frac{1}{E}\partial_e,
\end{eqnarray}
which if we make use of the unitary operators $U_j$ with $j\in{1,2,3}$ from the $(3+1)D$ flat spacetime on a cube lattice we can get our $\hat{W}$ operator as
\begin{eqnarray}
     \hat{W}&=&\mathbb{I}_4-\ii\epsilon\hat{H}+\mathcal{O}(\epsilon^2)\nonumber\\
     &=&\mathbb{I}_4-\ii m\epsilon(\sigma_3\otimes\mathbb{I}_2)-\epsilon(\sigma_1\otimes\sigma_1)\partial_1-\epsilon(\sigma_1\otimes\sigma_2)\frac{1}{C}\partial_c\nonumber\\&&-\epsilon(\sigma_1\otimes\sigma_2)\frac{1}{D}\partial_d-\epsilon(\sigma_1\otimes\sigma_3)\frac{1}{A}\partial_a-\epsilon(\sigma_1\otimes\sigma_3)\frac{1}{B}\partial_b\nonumber\\&&-\epsilon(\sigma_1\otimes\sigma_3)\frac{1}{E}\partial_e+\mathcal{O}(\epsilon^2)\nonumber\\
     &=&(\mathbb{I}_4-\ii m\epsilon(\sigma_3\otimes\mathbb{I}_2))\left(\mathbb{I}_4-\epsilon(\sigma_1\otimes\sigma_3)\frac{1}{E}\partial_e\right)\nonumber\\&&\times\left(\prod_{j\in\{d,c\}}\left(\mathbb{I}_4-\epsilon(\sigma_1\otimes\sigma_2)\frac{1}{J}\partial_j\right)\right)\nonumber\\
     &&\times\left(\prod_{j\in\{b,a\}}\left(\mathbb{I}_4-\epsilon(\sigma_1\otimes\sigma_3)\frac{1}{J}\partial_j\right)\right)\nonumber\\&&\times\left(\mathbb{I}_4-\epsilon(\sigma_1\otimes\sigma_1)\partial_1\right)+\mathcal{O}(\epsilon^2)\nonumber\\
     &=&\begin{pmatrix}\ee^{-\ii m\epsilon}\mathbb{I}_2&0\\0&\ee^{\ii m\epsilon}\mathbb{I}_2\end{pmatrix}(U_1\sigma_1\otimes U_3)\hat{S}_e(\sigma_1U_1^\dagger\otimes U_3^\dagger)\nonumber\\&&\times\left(\prod_{j\in\{d,c\}}\left((U_1\sigma_1\otimes U_2)\hat{S}_j(\sigma_1 U_1^\dagger\otimes U_2^\dagger)\right)\right)\nonumber\\&&\times\left(\prod_{j\in\{b,a\}}\left(U_1\sigma_1\otimes U_3)\hat{S}_j(\sigma_1 U_1^\dagger\otimes U_3^\dagger)\right)\right)\nonumber\\&&\times\left((U_1\sigma_1\otimes U_1)\hat{S}_1(\sigma_1 U_1^\dagger\otimes U_1^\dagger)\right).
\end{eqnarray}

Coupling this to an electromagnetic field would be trivial, as all that would be required is to include the operators from the cube walk that involve the $A_\mu$ terms as they have no effect on the shift operators.

\subsubsection{Rhombohedral Lattice (changing the $t$ step size) doesn't work}

It would be sensible at this point to think we could try a similar trick that was used in the triangular lattice to see if we can keep all the edges of our parallelepiped equal as in the case of a rhombohedral lattice. Unfortunately this fails to work, as when you follow the same procedure of moving the $\alpha$ off the $\partial_1$ term, there is no real solution for $\alpha$. This suggests that whereas before we could replace dilating in the $y-$dimension with a dilation in the time dimension, here we need to dilate in two dimensions - both $y$ and $z$. The time dimension can then only replace one of them. This suggests, at least under this construction, that a DQW on a rhombohedral lattice is not possible.

\section{Representation Invariance}

The previous section developed a DQW that works on a cubic lattice in a very particular representation - the standard Dirac representation. It is worth noting that this construction process does not require this choice of representation, it is in fact independent of the choice - although slightly different QWs will arise at the end of the construction. The first point to consider is the fact that any representation of the Dirac gamma matrices used in the Dirac equation can be related to each other by a unitary transformation. That is,
\begin{equation}
    \gamma'^\mu=\tilde{U}\gamma^\mu \tilde{U}^\dagger. \label{repchange}
\end{equation}
If we write the Hamiltonian (for arbitrary dimensions or lattice in free space) in terms of these gamma matrices instead we get
\begin{equation}
    \hat{H}=\tilde{U}\gamma^0\tilde{U}^\dagger m\epsilon-\sum_{j=1}^3\ii\tilde{U}\gamma^0\gamma^j\tilde{U}^\dagger\partial_j.
\end{equation}
The $\tilde{U}$ operators can easily be pulled out to the sides of our $\hat{W}$ operator by taking advantage of their unitary nature,
\begin{eqnarray}
    \hat{W}&=&\left(\mathbb{I}_4-\ii m\epsilon\tilde{U}\gamma^0\tilde{U}^\dagger\right)\prod_{j=1}^3\left(\mathbb{I}_4-\epsilon\tilde{U}\gamma^0\gamma^j\tilde{U}^\dagger\partial_j\right)+\mathcal{O}(\epsilon^2)\nonumber\\
    &=&\left(\tilde{U}\tilde{U}^\dagger-\ii m\epsilon\tilde{U}\gamma^0\tilde{U}^\dagger\right)\nonumber\\&&\times\prod_{j=1}^3\left(\tilde{U}\tilde{U}^\dagger-\epsilon\tilde{U}\gamma^0\gamma^j\tilde{U}^\dagger\partial_j\right)+\mathcal{O}(\epsilon^2)\nonumber\\
    &=&\tilde{U}\left(\mathbb{I}_4-\ii m\epsilon\gamma^0\right)\tilde{U}^\dagger\nonumber\\&&\times\prod_{j=1}^3\tilde{U}\left(\mathbb{I}_4-\epsilon\gamma^0\gamma^j\partial_j\right)\tilde{U}^\dagger+\mathcal{O}(\epsilon^2)\nonumber\\
    &=&\tilde{U}\left(\mathbb{I}_4-\ii m\epsilon\gamma^0\right)\left(\prod^3_{j=1}\left(\mathbb{I}_4-\epsilon\gamma^0\gamma^j\partial_j\right)\right)\tilde{U}^\dagger+\mathcal{O}(\epsilon^2).\nonumber\\
\end{eqnarray}
This is an arbitrary dimension, arbitrary lattice DQW with the standard Dirac representation, only with the unitary matrices wrapped around it. Thus a change of representation from the unitary transformation \ref{repchange} is dealt by doing the same unitary transformation to the walker operator $\hat{W}$,
\begin{equation}
    \hat{W'}=\tilde{U}\hat{W}\tilde{U}^\dagger,
\end{equation}
which in the case of the $(3+1)D$ on a cubic lattice this would then be,
\begin{eqnarray}
    \hat{W}&=&\tilde{U}\begin{pmatrix}
        \ee^{-\ii m\epsilon}\mathbb{I}_2&0\\
        0&\ee^{\ii m\epsilon}\mathbb{I}_2
    \end{pmatrix}\nonumber\\&&\times\left(\prod^3_{j=1}\left((U_1\sigma_1\otimes U_j)\hat{S}_j(\sigma_1U_1^\dagger\otimes U_j^\dagger)\right)\right)\tilde{U}^\dagger.\nonumber\\
\end{eqnarray}
So as long as this construction method can derive a DQW on a particular lattice in the Dirac representation, it can derive a DQW in any representation.

\section{Conclusion}

We have introduced a new systematic method to construct DQWs on regular lattices of arbitrary dimensions. This method is superior to a trial-and-error method because (i) it shows unambiguously if a DQW can be constructed on a given lattice (ii) it delivers automatically the coefficients of the DQW (iii) it becomes necessary to use a systematic approach if one wants to deal with lattices of dimensions higher than $2$. We have presented the method in a pedagogical manner on two relatively simple cases, a $3D$ DQW on the cubic lattice coupled to an EM field and a $2D$ DQW on the triangular lattice. 
The free DQW on the cubic $3D$ lattice is well-known \cite{arrighi2014dirac} but its coupling to an arbitrary electromagnetic field is not. The $2D$ DQW on the triangle lattice has been presented in \cite{jay2018dirac} but without directional derivatives. The new presentation given here shows that the coefficients and scalings of this walk are the only ones that work on this lattice. Note that directional derivatives have already been used in 
writing a DQW on triangular lattice \cite{arrighi2018dirac}, but the spinor of that walk does not live on the vertices of the lattice, but on its edges. We have finally constructed a new DQW on the parallelepiped lattice and produced as a negative example in the equilateral rhombohedral lattice, where the method shows no DQW can be constructed. 

Let us conclude by mentioning possible extensions of this work. One should first extend the procedure to include arbitrary Yang-Mills and gravitational fields, thus producing QWs which model the Dirac equation in curved spacetime coupled to arbitrary gauge fields. Also, DQWs with spinors living on the edges of a graph have been introduced recently in \cite{arrighi2018dirac, arrighi2018curved}. The method we present in this article should thus be extended to take this possibility into account, as well as wave-functions with a higher number of components \cite{arrighi2016quantum} than the number of spinor components of irreducible representations of the Lorentz group. One wonders for example if these extensions would make it possible to define DQWs on lattices where the present method fails, as the rombohedral lattice. Finally, studying systematically how to produce DQWs on both regular and non regular graphs is of paramount importance.

\bibliographystyle{apsrev4-1}
\bibliography{Biblio.bib}

\appendix
\section{Appendix}
\subsection{A pedagogical walkthrough of the method}
The following appendix is designed to be an easy to follow explanation of the Hamiltonian process applied to hypercube lattices starting at the most simple case of a $(1+1)D$ Dirac equation in free space, and building up to the more complicated cases as the dimensions increase.
Tools we will need to consider are the following expansions:
\begin{eqnarray}
	\cos(\epsilon A)&=&\mathbb{I}_N+\mathcal{O}(\epsilon^2)\label{cos},\\
    \sin(\epsilon A)&=&\epsilon A + \mathcal{O}(\epsilon^2)\label{sin},\\
    \ee^{\epsilon A} &=&\mathbb{I}_N+\epsilon A +\mathcal{O}(\epsilon^2)\label{exp},
\end{eqnarray}
while the shift operators, at least in the lower dimensions of $(1+1)D$ and $(2+1)D$ spacetime, used in the walks can be expressed as an exponential of differential operators like so,
\begin{eqnarray}
\hat{S}_1\Psi(x,t)&=&\begin{pmatrix}
								\psi_L(x+\epsilon,t)\\
                                \psi_R(x-\epsilon,t)
					 \end{pmatrix}\nonumber\\
					 &=&\begin{pmatrix}
					    \psi_L(x,t)+\epsilon\partial_1\psi_L(x,t)+\mathcal{O}(\epsilon^2)\\
					    \psi_R(x,t)-\epsilon\partial_1\psi_R(x,t)+\mathcal{O}(\epsilon^2)
					 \end{pmatrix}\nonumber\\
                 &=&\begin{pmatrix}
                 		\psi_L(x,t)\\
                        \psi_R(x,t)
                    \end{pmatrix}+\epsilon\sigma_3\partial_1\begin{pmatrix}
                 												\psi_L(x,t)\\
                       											\psi_R(x,t)
                    										\end{pmatrix}+\mathcal{O}(\epsilon^2)\nonumber\\
                 &=&\left(\mathbb{I}_2+\epsilon\sigma_3\partial_1\right)\begin{pmatrix}
                 												\psi_L(x,t)\\
                       											\psi_R(x,t)
                    										\end{pmatrix}+\mathcal{O}(\epsilon^2)\nonumber\\
              &=&\ee^{\epsilon\sigma_3\partial_1}\Psi(x,t).
\end{eqnarray}
This works for the other spatial directions so that in general the Shift Operator (acting on a 2-vector wavefunction $\Psi$) is
\begin{equation}\label{lowDimShiftDef}
	\hat{S}_i=\ee^{\epsilon\sigma_3\partial_i}.
\end{equation}
We now have the tools to start tackling the simplest individual cases.
\subsubsection{$(1+1)D$ flat spacetime in free space}
 The simplest case is the $(1+1)D$ free flat spacetime version of the Dirac Equation:
 \begin{equation}
 		(\ii\gamma^\mu\partial_\mu-m)\Psi=0,
 \end{equation}
 where we will use the standard $2\times 2$ Gamma matrices of $\gamma^0=\sigma_1$ and $\gamma^1=\ii\sigma_2$.  First we need to rearrange this into Hamiltonian form to identify the Hamiltonian $\hat{H}$:
 \begin{equation}
 	\hat{H}=\ii\sigma_3\partial_1+\sigma_1m.
 \end{equation}
So therefore the operator $\hat{W}$ for our DTQW will take the form
\begin{eqnarray}
	\hat{W}&=&\mathbb{I}_2-\ii\epsilon\hat{H}+\mathcal{O}(\epsilon^2)\nonumber\\
    &=&\mathbb{I}_2+\epsilon\sigma_3\partial_1-\ii m\epsilon\sigma_1+\mathcal{O}(\epsilon^2)\nonumber\\
    &=&(\mathbb{I}_2-\ii m\epsilon\sigma_1)(\mathbb{I}_2+\epsilon\sigma_3\partial_1)+\mathcal{O}(\epsilon^2)\nonumber\\
    &=&\begin{pmatrix}
    1&-\ii m\epsilon\\
    -\ii m\epsilon&1
    \end{pmatrix}e^{\epsilon\sigma_3\partial_1}+\mathcal{O}(\epsilon^2)\nonumber\\
    &=&\begin{pmatrix}
    \cos(m\epsilon)&-\ii\sin(m\epsilon)\\
    -\ii\sin(m\epsilon)&\cos(m\epsilon)
    \end{pmatrix}\hat{S}_1.
\end{eqnarray}
Here step 4 has made use of definition \ref{lowDimShiftDef}, while step 5 has used the expansions \ref{cos} and \ref{sin}. And so the DTQW that reaches the $(1+1)D$ free flat spacetime Dirac Equation in the continuous limit would be 
\begin{equation}
	\Psi(x,t+\epsilon)=\begin{pmatrix}
    \cos(m\epsilon)&-\ii\sin(m\epsilon)\\
    -\ii\sin(m\epsilon)&\cos(m\epsilon)
    \end{pmatrix}\begin{pmatrix}
								\psi_L(x+\epsilon,t)\\
                                \psi_R(x-\epsilon,t)
					 \end{pmatrix},
\end{equation}
which agrees with previous works results \cite{chandrashekar2010relationship}.
\subsubsection{$(1+1)D$ flat spacetime coupled to an electric field}
The Dirac Equation now coupled to an electric field takes the form
\begin{equation}
	(\ii\gamma^\mu D_\mu-m)\Psi=0,
\end{equation}
where 
\begin{equation}
	D_\mu =\partial_\mu-\ii A_\mu.
\end{equation}
Employing the same technique we get the Hamiltonian by rearranging the equation:
\begin{equation}
	\hat{H}=-A_0+\sigma_3A_1+\sigma_1m+\ii\sigma_3\partial_1.
\end{equation}
And so the operator $\hat{W}$ takes the form
\begin{eqnarray}
	\hat{W}&=&\mathbb{I}_2-\ii\epsilon\hat{H}+\mathcal{O}(\epsilon^2)\nonumber\\
     &=&\mathbb{I}_2+\ii\epsilon A_0-\ii\epsilon\sigma_3A_1-\ii\epsilon\sigma_1m+\epsilon\sigma_3\partial_1+\mathcal{O}(\epsilon^2)\nonumber\\
    &=&(\mathbb{I}_2+\ii\epsilon A_0)(\mathbb{I}_2-\ii\epsilon\sigma_3 A_1)(\mathbb{I}_2-\ii\epsilon\sigma_1m)(\mathbb{I}_2+\epsilon\sigma_3\partial_1)\nonumber\\&&+\mathcal{O}(\epsilon^2).\nonumber
\end{eqnarray}
Now clearly the last two brackets in this expression are identical to the free space case from before. For the two new expressions we make use of expansion \ref{exp} and get:
\begin{eqnarray}
    \Hat{W}&=&e^{\ii\epsilon A_0}\begin{pmatrix}
    1-\ii\epsilon A_1&0\\
    0&1+\ii\epsilon A_1
    \end{pmatrix}\nonumber\\&&\times\begin{pmatrix}
    \cos(m\epsilon)&-\ii\sin(m\epsilon)\\
    -\ii\sin(m\epsilon)&\cos(m\epsilon)
    \end{pmatrix}\hat{S}_1+\mathcal{O}(\epsilon^2)\nonumber\\
    &=&\ee^{\ii\epsilon A_0}\begin{pmatrix}
    \ee^{-\ii\epsilon A_1}&0\\
    0&\ee^{\ii\epsilon A_1}
    \end{pmatrix}\begin{pmatrix}
    \cos(m\epsilon)&-\ii\sin(m\epsilon)\\
    -\ii\sin(m\epsilon)&\cos(m\epsilon)
    \end{pmatrix}\hat{S}_1\nonumber\\
    &=&\ee^{\ii\epsilon A_0}\begin{pmatrix}
    \ee^{-\ii\epsilon A_1}\cos(m\epsilon) & \ee^{-\ii(\epsilon A_1+\frac{\pi}{2})}\sin(m\epsilon)\\
    -\ee^{\ii(\epsilon A_1+\frac{\pi}{2})}\sin(m\epsilon)&\ee^{\ii\epsilon A_1}\cos(m\epsilon)
    \end{pmatrix}\hat{S}_1.\nonumber\\
\end{eqnarray}
Which matches previous work stating you need the more general unitary coin with all four angles set to non-zero to reach in the limit a Dirac equation coupled to an electric field \cite{di2014quantum}.

\subsubsection{$(2+1)D$ flat spacetime in free space}
Going up one dimension, the Hamiltonian form of the Dirac Equation contains an extra differential term that requires dealing with. As such we will need another gamma matrix. Fortunately in $(2+1)D$ spacetime, 2x2 matrices still suffice and we shall use the standard $\gamma^2=\ii\sigma_3$. With this in mind, the Hamiltonian can be determined as
\begin{equation}
	\hat{H}=\sigma_1 m+\ii\sigma_3\partial_1-\ii\sigma_2\partial_2.\label{2+1FreeH}
\end{equation}
The operator $\hat{W}$ will now take the form of
\begin{eqnarray}
	\hat{W}&=&\mathbb{I}_2-\ii\epsilon\hat{H}+\mathcal{O}(\epsilon^2)\nonumber\\
    &=&\mathbb{I}_2-\ii m\epsilon\sigma_1+\epsilon\sigma_3\partial_1-\epsilon\sigma_2\partial_2+\mathcal{O}(\epsilon^2)\nonumber\\
    &=&(\mathbb{I}_2-\ii m\epsilon\sigma_1)(\mathbb{I}_2-\epsilon\sigma_2\partial_2)(\mathbb{I}_2+\epsilon\sigma_3\partial_1)+\mathcal{O}(\epsilon^2)\nonumber\\
    &=&\begin{pmatrix}
    \cos(m\epsilon)&-\ii\sin(m\epsilon)\\
    -\ii\sin(m\epsilon)&\cos(m\epsilon)
    \end{pmatrix}(\mathbb{I}_2-\epsilon\sigma_2\partial_2)\hat{S}_1+\mathcal{O}(\epsilon^2).\nonumber
\end{eqnarray}
This is when we run into trouble with the partial derivative in the $y-$direction. Where we really want a $\sigma_3$ to be sitting so that we can bring in the second shift operator $\hat{S}_2$ we have instead a $-\sigma_2$. However if we can find a unitary matrix $U$ such that
\begin{equation}\label{UDef}
	U\sigma_3U^\dagger=-\sigma_2,
\end{equation}
then we can continue as
\begin{eqnarray}
    \Hat{W}&=&\begin{pmatrix}
    \cos(m\epsilon)&-\ii\sin(m\epsilon)\\
    -\ii\sin(m\epsilon)&\cos(m\epsilon)
    \end{pmatrix}(\mathbb{I}_2-\epsilon\sigma_2\partial_2)\hat{S}_1+\mathcal{O}(\epsilon^2)\nonumber\\
    &=&\begin{pmatrix}
    \cos(m\epsilon)&-\ii\sin(m\epsilon)\\
    -\ii\sin(m\epsilon)&\cos(m\epsilon)
    \end{pmatrix}(UU^\dagger+\epsilon U\sigma_3U^\dagger\partial_2)\hat{S}_1\nonumber\\&&+\mathcal{O}(\epsilon^2)\nonumber\\
    &=&\begin{pmatrix}
    \cos(m\epsilon)&-\ii\sin(m\epsilon)\\
    -\ii\sin(m\epsilon)&\cos(m\epsilon)
    \end{pmatrix}U(\mathbb{I}_2+\epsilon \sigma_3\partial_2)U^\dagger\hat{S}_1\nonumber\\&&+\mathcal{O}(\epsilon^2)\nonumber\\
    &=&\begin{pmatrix}
    \cos(m\epsilon)&-\ii\sin(m\epsilon)\\
    -\ii\sin(m\epsilon)&\cos(m\epsilon)
    \end{pmatrix}U\hat{S}_2U^\dagger\hat{S}_1.
\end{eqnarray}
So as long as we can solve equation \ref{UDef} then we have our DQW. We shall see that $U$ is not unique. If we define $U$ as 
\begin{equation}
    U=\begin{pmatrix}
        a&b\\
        c&d\\
    \end{pmatrix},
\end{equation}
we then have two requirements. Namely equation \ref{UDef} which leads to
\begin{equation}
    \begin{pmatrix}
        |a|^2-|b|^2&a\bar{c}-b\bar{d}\\
        c\bar{a}-d\bar{b}&|c|^2-|d|^2\\
    \end{pmatrix}=\begin{pmatrix}
        0&i\\
        -i&0\\
    \end{pmatrix},
\end{equation}
and the fact that $U$ is unitary so that
\begin{equation}
     \begin{pmatrix}
        |a|^2+|b|^2&a\bar{c}+b\bar{d}\\
        c\bar{a}+d\bar{b}&|c|^2+|d|^2\\
    \end{pmatrix}=\begin{pmatrix}
        1&0\\
        0&1\\
    \end{pmatrix}.
\end{equation}
This ultimately leads to the relationships
\begin{eqnarray}
    |a|^2=|b|^2=|c|^2=|d|^2&=&\frac{1}{2},\\
    c\bar{a}=b\bar{d}&=&-\frac{\ii}{2}.
\end{eqnarray}
There are therefore two degrees of freedom in the choice of $U$, so if we use the first relationship to choose $a$ and $d$ to be $\frac{1}{\sqrt{2}}$ then the second relationship will fix $b$ and $c$ for us and $U$ is
\begin{equation}
U=\begin{pmatrix}
\frac{1}{\sqrt{2}}&-\frac{\ii}{\sqrt{2}}\\
-\frac{\ii}{\sqrt{2}}&\frac{1}{\sqrt{2}}
\end{pmatrix}.
\end{equation}
\subsubsection{$(2+1)D$ flat spacetime coupled to an electromagnetic field}
This time the Hamiltonian will have more terms, however none of the extra terms will present any challenge to us.
\begin{equation}
	\hat{H}=-A_0+\ii\sigma_3\partial_1+\sigma_3A_1-\ii\sigma_2\partial_2-\sigma_2A_3+\sigma_1m.
\end{equation}
The operator $\hat{W}$ simply becomes
\begin{eqnarray}
	\hat{W}&=&\mathbb{I}_2-\ii\epsilon\hat{H}+\mathcal{O}(\epsilon^2)\nonumber\\
    &=&\mathbb{I}_2+\ii\epsilon A_0+\epsilon\sigma_3\partial_1-\ii\epsilon\sigma_3A_1-\epsilon\sigma_2\partial_2+\ii\epsilon\sigma_2A_2\nonumber\\&&-\ii\epsilon\sigma_1 m+\mathcal{O}(\epsilon^2)\nonumber\\
    &=&(\mathbb{I}_2+\ii\epsilon A_0)(\mathbb{I}_2-\ii\epsilon\sigma_3A_1)(\mathbb{I}_2+\ii\epsilon\sigma_2A_2)(\mathbb{I}_2-\ii\epsilon\sigma_1 m)\nonumber\\&&\times(\mathbb{I}_2-\epsilon\sigma_2\partial_2)(\mathbb{I}_2+\epsilon\sigma_3\partial_1)+\mathcal{O}(\epsilon^2)\nonumber\\
    &=&\ee^{\ii\epsilon A_0}\begin{pmatrix}
    						\ee^{-\ii\epsilon A_1} & 0\\
                            0 & \ee^{\ii\epsilon A_1}
					    \end{pmatrix}\begin{pmatrix}
					    				1 & \epsilon A_2\\
                                        -\epsilon A_2 & 1
					    			 \end{pmatrix}\nonumber\\&&\times\begin{pmatrix}
					    			 									\cos(m\epsilon) & -\ii\sin(m\epsilon)\\
                                                                        -\ii\sin(m\epsilon) & \cos(m\epsilon)
					    			 								\end{pmatrix}U\hat{S}_2U^\dagger\hat{S}_1+\mathcal{O}(\epsilon^2)\nonumber\\
    &=&\ee^{\ii\epsilon A_0}\begin{pmatrix}
    						\ee^{-\ii\epsilon A_1} & 0\\
                            0 & \ee^{\ii\epsilon A_1}
					    \end{pmatrix}\begin{pmatrix}
					    				\cos(\epsilon A_2) & \sin(\epsilon A_2)\\
                                        -\sin(\epsilon A_2) & \cos(\epsilon A_2)
					    			 \end{pmatrix}\nonumber\\&&\times\begin{pmatrix}
					    			 									\cos(m\epsilon) & -\ii\sin(m\epsilon)\\
                                                                        -\ii\sin(m\epsilon) & \cos(m\epsilon)
					    			 								\end{pmatrix}U\hat{S}_2U^\dagger\hat{S}_1.
\end{eqnarray}

Note by now we can start to see a pattern in the method. Adding extra differential terms to our Dirac Equation simply means adding shift operators usually wrapped in a unitary transformation. Adding non-differential terms to our Dirac equation is much simpler in that all it does is add unitary operators to the front of the walk operator.

\subsubsection{$(3+1)D$ Flat Spacetime in free spacetime}

Increasing to the full $(3+1)D$ spacetime, the Dirac Equation can no longer obey the Clifford Algebra for the Gamma Matrices is we remain in a 2-component spinor space. The Gamma matrices must instead now be 4x4 matrices and the wavefunction $\Psi$, a 4-vector.
\begin{equation}
	\Psi=\begin{pmatrix}
		\psi_1\\
        \psi_2\\
        \psi_3\\
        \psi_4
	\end{pmatrix}
\end{equation}
Here $\{\psi_1,\psi_2\}$ are the spin-up/spin-down components of a particle and $\{\psi_3,\psi_4\}$ represent an antiparticle. Our new shift operators must now tackle four components and look like:
\begin{eqnarray}
	\hat{S}_1\Psi(x,y,z,t)&=&\begin{pmatrix}
	\psi_1(x+\epsilon,y,z,t)\\
    \psi_2(x-\epsilon,y,z,t)\\
    \psi_3(x-\epsilon,y,z,t)\\
    \psi_4(x+\epsilon,y,z,t)
	\end{pmatrix}\nonumber\\
	&=&\begin{pmatrix}
	\psi_1(x,y,z,t)+\epsilon\partial_1\psi_1(x,y,z,t)+\mathcal{O}(\epsilon^2)\\
    \psi_2(x,y,z,t)-\epsilon\partial_1\psi_2(x,y,z,t)+\mathcal{O}(\epsilon^2)\\
    \psi_3(x,y,z,t)-\epsilon\partial_1\psi_3(x,y,z,t)+\mathcal{O}(\epsilon^2)\\
    \psi_4(x,y,z,t)+\epsilon\partial_1\psi_4(x,y,z,t)+\mathcal{O}(\epsilon^2)
	\end{pmatrix}\nonumber\\
	&=&\begin{pmatrix}
	\psi_1(x,y,z,t)\\
    \psi_2(x,y,z,t)\\
    \psi_3(x,y,z,t)\\
    \psi_4(x,y,z,t)
	\end{pmatrix}\nonumber\\&&+\epsilon(\sigma_3\otimes\sigma_3)\partial_1\begin{pmatrix}
	\psi_1(x,y,z,t)\\
    \psi_2(x,y,z,t)\\
    \psi_3(x,y,z,t)\\
    \psi_4(x,y,z,t)
	\end{pmatrix}+\mathcal{O}(\epsilon^2)\nonumber\\
    &=&(\mathbb{I}_4+\epsilon(\sigma_3\otimes\sigma_3)\partial_1)\begin{pmatrix}
	\psi_1(x,y,z,t)\\
    \psi_2(x,y,z,t)\\
    \psi_3(x,y,z,t)\\
    \psi_4(x,y,z,t)
	\end{pmatrix}\nonumber\\
    &=&\ee^{\epsilon(\sigma_3\otimes\sigma_3)\partial_1}\Psi(x,y,z,t).
\end{eqnarray}
This works for the other spatial directions so that in general the Shift Operator (acting on a 4-vector wavefunction $\Psi$) is
\begin{equation}
	\hat{S}_i=\ee^{\epsilon(\sigma_3\otimes\sigma_3)\partial_i}.
\end{equation}

For our 4x4 Gamma matrices we shall use the standard representation choice of
\begin{eqnarray}
	\gamma^0&=&\begin{pmatrix}
	1&0&0&0\\
    0&1&0&0\\
    0&0&-1&0\\
    0&0&0&-1
	\end{pmatrix}
	=\begin{pmatrix}
	\mathbb{I}_2&0\\
    0&-\mathbb{I}_2
	\end{pmatrix}=\sigma_3\otimes\mathbb{I}_2,\\
    \gamma^1&=&\begin{pmatrix}
    	0&0&0&1\\
        0&0&1&0\\
        0&-1&0&0\\
        -1&0&0&0
    \end{pmatrix}=\begin{pmatrix}
    0&\sigma_1\\
    -\sigma_1&0
    \end{pmatrix}=\ii\sigma_2\otimes\sigma_1,\\
    \gamma^2&=&\begin{pmatrix}
    0&0&0&-\ii\\
    0&0&\ii&0\\
    0&\ii&0&0\\
    -\ii&0&0&0
    \end{pmatrix}=\begin{pmatrix}
    0&\sigma_2\\
    -\sigma_2&0
    \end{pmatrix}=\ii\sigma_2\otimes\sigma_2,\\
    \gamma^3&=&\begin{pmatrix}
    0&0&1&0\\
    0&0&0&-1\\
    -1&0&0&0\\
    0&1&0&0
    \end{pmatrix}=\begin{pmatrix}
    0&\sigma_3\\
    -\sigma_3&0
    \end{pmatrix}=\ii\sigma_2\otimes\sigma_3.
\end{eqnarray}
In free space the Hamiltonian becomes
\begin{eqnarray}
\hat{H}=(\sigma_3\otimes\mathbb{I}_2)m-\sum_{j=1}^3\ii(\sigma_1\otimes\sigma_j)\partial_j.
\end{eqnarray}
Which leads to a $\hat{W}$ operator of the form:
\begin{eqnarray}
	\hat{W}&=&\mathbb{I}_4-\ii\epsilon\hat{H}+\mathcal{O}(\epsilon^2)\nonumber\\
    &=&\mathbb{I}_4-\ii m\epsilon(\sigma_3\otimes\mathbb{I}_2)-\epsilon(\sigma_1\otimes\sigma_j)\partial_j+\mathcal{O}(\epsilon^2)\nonumber\\
    &=&(\mathbb{I}_4-\ii m\epsilon(\sigma_3\otimes\mathbb{I}_2))\prod^3_{j=1}\left(\mathbb{I}_4-\epsilon(\sigma_1\otimes\sigma_j)\partial_j\right)+\mathcal{O}(\epsilon^2).\nonumber
\end{eqnarray}
If we follow the pattern from before we realize the mass expression is not going to be difficult to deal with, however the partial derivatives we will need to find a way to use a unitary transformation to change those Pauli matrices to $\sigma_3$. Taking a similar trick to before, we now look for three different unitary matrices that satisfy
\begin{equation}
U_j\sigma_3U_j^\dagger=\sigma_j.
\end{equation}
If we make use of the mixed-product property of Kronecker products
\begin{equation}
    (A\otimes B)(C\otimes D)=(AC)\otimes(BD),
\end{equation}
then we can derive that
\begin{eqnarray}
    \hat{W}&=&(\mathbb{I}_4-\ii m\epsilon(\sigma_3\otimes\mathbb{I}_2))\nonumber\\&&\times\prod^3_{j=1}\left(\mathbb{I}_2\otimes\mathbb{I}_2-\epsilon(\sigma_1\otimes\sigma_j)\partial_j\right)+\mathcal{O}(\epsilon^2)\nonumber\\
    &=&(\mathbb{I}_4-\ii m\epsilon(\sigma_3\otimes\mathbb{I}_2))\nonumber\\&&\times\prod^3_{j=1}\left(U_1U_1^\dagger\otimes U_jU_j^\dagger-\epsilon(U_1\sigma_3U_1^\dagger\otimes U_j\sigma_3U_j^\dagger)\partial_j\right)\nonumber\\&&+\mathcal{O}(\epsilon^2)\nonumber\\
    &=&(\mathbb{I}_4-\ii m\epsilon(\sigma_3\otimes\mathbb{I}_2))\nonumber\\&&\times\prod^3_{j=1}\left((U_1\otimes U_j)(\mathbb{I}_2\otimes\mathbb{I}_2-\epsilon(\sigma_3\otimes \sigma_3)\partial_j)(U_1^\dagger\otimes U_j^\dagger)\right)\nonumber\\&&+\mathcal{O}(\epsilon^2)\nonumber\\
    &=&(\mathbb{I}_4-\ii m\epsilon(\sigma_3\otimes\mathbb{I}_2))\nonumber\\&&\times\prod^3_{j=1}\bigg((U_1\otimes U_j)(\sigma_1\sigma_1\otimes\mathbb{I}_2+\epsilon(\sigma_1\sigma_3\sigma_1\otimes \sigma_3)\partial_j)\nonumber\\&&\times(U_1^\dagger\otimes U_j^\dagger)\bigg)+\mathcal{O}(\epsilon^2)\nonumber\\
    &=&(\mathbb{I}_4-\ii m\epsilon(\sigma_3\otimes\mathbb{I}_2))\nonumber\\&&\times\prod^3_{j=1}\bigg((U_1\sigma_1\otimes U_j)(\mathbb{I}_4+\epsilon(\sigma_3\otimes \sigma_3)\partial_j)(\sigma_1 U_1^\dagger\otimes U_j^\dagger)\bigg)\nonumber\\&&+\mathcal{O}(\epsilon^2)\nonumber\\
    &=&\begin{pmatrix}
    \ee^{-\ii m\epsilon}\mathbb{I}_2&0\\
    0&\ee^{\ii m\epsilon}\mathbb{I}_2
    \end{pmatrix}\prod^3_{j=1}\left((U_1\sigma_1\otimes U_j)\hat{S}_j(\sigma_1 U_1^\dagger\otimes U_j^\dagger)\right).\nonumber\\
\end{eqnarray}
Like before in the $(2+1)D$ case these are not unique. Similar arguments can determine a couple of degrees of freedom and the choices we shall make here are the unitary matrices
\begin{eqnarray}
U_1&=&\begin{pmatrix}
\frac{1}{\sqrt{2}}&\frac{1}{\sqrt{2}}\\
\frac{1}{\sqrt{2}}&-\frac{1}{\sqrt{2}}
\end{pmatrix},\nonumber\\
U_2&=&\begin{pmatrix}
\frac{1}{\sqrt{2}}&\frac{1}{\sqrt{2}}\\
\frac{i}{\sqrt{2}}&-\frac{i}{\sqrt{2}}
\end{pmatrix},\nonumber\\
U_3&=&\mathbb{I}_2.
\end{eqnarray}
\subsubsection{$(3+1)D$ flat spacetime coupled to an electromagnetic field}
The Dirac equation coupled to an Electromagnetic field in $(3+1)D$ spacetime is shown in the main text, however as we can expect by now the only addition it will have over the free space version is an extra unitary operator for each extra term there is in the Dirac equation. In other words, three more operators, one for each $A_\mu$ term that appears in the equation
\begin{equation}
    (\ii\gamma^\mu D_\mu-m)\Psi=0,\quad D_\mu=\partial_\mu-iA_\mu.
\end{equation}

\end{document}